\documentclass[aps,12pt,final,notitlepage,oneside,onecolumn,nobibnotes,nofootinbib,superscriptaddress,noshowpacs,centertags]{revtex4}

\usepackage{amsmath,amssymb,epsfig,placeins,subfigure,wrapfig,indentfirst,makeidx}
\usepackage[cp1251]{inputenc}
\usepackage[english]{babel}
\usepackage{amsfonts,amssymb}
\usepackage{mathrsfs}
\usepackage{dsfont}

\begin{document}

\title{Model Predictions of the Results of Interferometric Observations for Stars under Conditions of
Strong Gravitational Scattering by Black Holes and Wormholes}

\author{Alexander Shatskiy}
\affiliation{\,\, shatskiyalex@gmail.com}

\author{Yu. Yu. Kovalev}
\affiliation{Astro Space Center, Lebedev Physical Institute,
Russian Academy of Sciences, Profsoyuznaya ul. 84/32, Moscow,
117997 Russia}

\author{I. D. Novikov}
\affiliation{Astro Space Center, Lebedev Physical Institute,
Russian Academy of Sciences, Profsoyuznaya ul. 84/32, Moscow,
117997 Russia} \affiliation{The Nielse Bohr International Academy,
The Nielse Bohr Institute, Blegdamsvej 17, DK-2100 Copenhagen,
Denmark}

\begin{abstract}
The characteristic and distinctive features of the visibility
amplitude of interferometric observations for compact objects like
stars in the immediate vicinity of the central black hole in our
Galaxy are considered. These features are associated with the
specifics of strong gravitational scattering of point sources by
black holes, wormholes, or black–white holes. The revealed
features will help to determine the most important topological
characteristics of the central object in our Galaxy: whether this
object possesses the properties of only a black hole or also has
characteristics unique to wormholes or black–white holes. These
studies can be used to interpret the results of optical, infrared,
and radio interferometric observations.
\end{abstract}

\maketitle

\section{INTRODUCTION}
\label{s1} Observations of many objects in the Universe with
single telescopes, even space ones, do not allow their structure
to be investigated, because their angular sizes are small. At the
same time, the angular resolution of present day long baseline
optical, infrared, and radio interferometers (the latter are the
so called very long baseline interferometers, VLBI) approaches ten
microarcseconds\footnote{1 microarcsecond ${(mas) \approx 4.8\cdot
10^{-12}}$ rad.} see, for example, \citep[see, for
example,][]{VLBIbook, KM, Q_rev, Q_space, Lu, RA}). However, an
interferometer records not the image of a compact object itself
but its complex Fourier transform $V$ or, as it is also called the
“visibility function” (see \citep[]{VLBIbook} for more details):
\begin{eqnarray}
V({\rm u,v}) = \int\limits\int\limits I(x,y) \exp\left[ -2\pi
i({\rm xu+yv})/\lambda \right] \, dx\, dy \label{furye1}
\end{eqnarray}
Here, ${I}$ is the intensity of the image; ${(x,y)}$ are its
angular coordinates; ${\rm (u,v)}$  are the coordinates of the
interferometer baseline projection onto the ${(x,y)}$ plane; and
$\lambda$ is the wavelength at which the fringe pattern is
observed. The complex expression (\ref{furye1}) has an amplitude
and a phase (which is defined by the phase difference between the
signal arrivals at the interferometer’s telescopes). The inverse
Fourier transform should be performed to synthesize the image from
the Fourier transform:
\begin{eqnarray}
I(x,y) = \int\limits\int\limits V({\rm u,v}) \exp\left[ 2\pi
i({\rm xu+yv})/\lambda \right] \, {\rm du\, dv} \label{furye2}
\end{eqnarray}
As is well known (see, e.g., \citep[]{VLBIbook}), the phase is a
no less important function than the amplitude when the image is
reconstructed from the Fourier transform. Moreover, it is
virtually impossible to properly reconstruct the image without
knowing the phase. However, not all interferometers are able to
measure the visibility function phase. In addition, real
interferometric observations do not give a filled ${\rm (u,v)}$
plane, which makes it difficult to reconstruct the image of an
object and limits the dynamic range of the resulting map. The
method of directly modeling and comparing the results of
visibility function amplitude (or Correlated Flux Density -- CFD)
measurements for the compact object being investigated and the
model by the $\chi^2$ minimization technique can be used to solve
these problems. In this paper, we will attempt to find the
characteristic features in the CFD for compact objects that can be
associated with strong gravitational scattering of stars in the
field of a black hole or wormhole. A similar approach is applied
by the group of the Event Horizon Telescope (see \cite{crescent,
SILHOUETTE2009}). We will note at once that the stars we consider
are bright, point, and compact objects that cannot be resolved.
However, the stars can produce a system of fairly bright point
images through strong gravitational scattering. This system will
make a characteristic contribution to the CFD formation, because
the sizes of this system of images will be large enough for its
angular resolution with an interferometer but still insufficient
for its observation with a single telescope. A thin shining ring
that, according to theoretical predictions, must be seen around a
Schwarzschild black hole (the apparent diameter of this ring in
linear units is ${3\sqrt{3}r_g/2}$, where $r_g$ is the
gravitational radius) can be a typical example of such a system of
images from stars close to a black hole.

As the main object of investigation we will choose the object at
the center of our Milky Way, a massive black hole with a mass
${\approx 4.3\cdot 10^6 M_\odot}$ and a gravitational radius
${\approx 13\cdot 10^6}$ km (see, e.g., \citep[]{S2}).

The supermassive black hole in the quasar M87 with a mass of
${\approx 3.4\cdot 10^9 M_\odot}$ (see, e.g.,
\citep[]{SILHOUETTE2009}) could be yet another possible candidate
for an object of observation, because the angular size of the
horizon radius for this black hole turns out to be sufficient
(${r_g/r\approx 4.7\,\mu as \approx 2.3\cdot 10^{-11}}$ rad).
However, with this quasar being too far away from us (${\approx
16}$ Mpc), we cannot count on the detection of individual stars,
including bright radio pulsars, with interferometers.

Since the effects being investigated are expected to be observed
at the sensitivity limit of interferometers, for solar type stars
it makes sense to consider infrared and optical observations,
i.e., for ${\lambda\sim 1\mu m}$. It is in this range that the
brightness of stars is at a maximum. However, an assumption about
the pattern of the radiation spectrum for stars and background
should also be made in the optical and infrared ranges (as, for
example, was done in \citep[]{SILHOUETTE2009}). The point is that
the relative width of the detected spectral band ${(\Delta\lambda
/\lambda)}$ is great in the optical and infrared ranges. For
example, we can assume a flat spectrum or choose a spectrum
typical of a solar type star.

In the radio band, VLBI systems record relatively narrow band
signals, and there is no need to make any assumptions about the
spectrum shape. However, the brightness of ordinary stars is
insufficient, and pulsars can be considered in principle as the
bright and point sources needed for us. Our subsequent
calculations will be performed under the assumption of a
monochromatic spectrum.

\section{BRIGHTNESS VARIATIONS DURING THE PROPAGATION OF A THIN LIGHT BEAM}
\label{s2} To avoid misunderstandings, we will note at once: by
strong gravitational scattering we will mean not the brightness
amplification in sources (as is usually meant when gravitational
scattering is considered) but, on the contrary, the attenuation of
their brightness through the scattering of light by the
gravitational lens. Therefore, we will not be interested in the
standard formulas and conclusions of gravitational scattering,
where the observer is assumed to be almost in the focal plane of
the lens. Since these are highly unlikely joint conditions for the
source, the gravitational center, and the observer, quite the
reverse is true in our case: the observer is far from the focal
plane of the lens. Indeed, only a very small fraction of celestial
sources (from their total number) turn out to be gravitationally
lensed by some bodies in the Universe for an observer on Earth.

\begin{figure}
\includegraphics[width=0.8\textwidth]{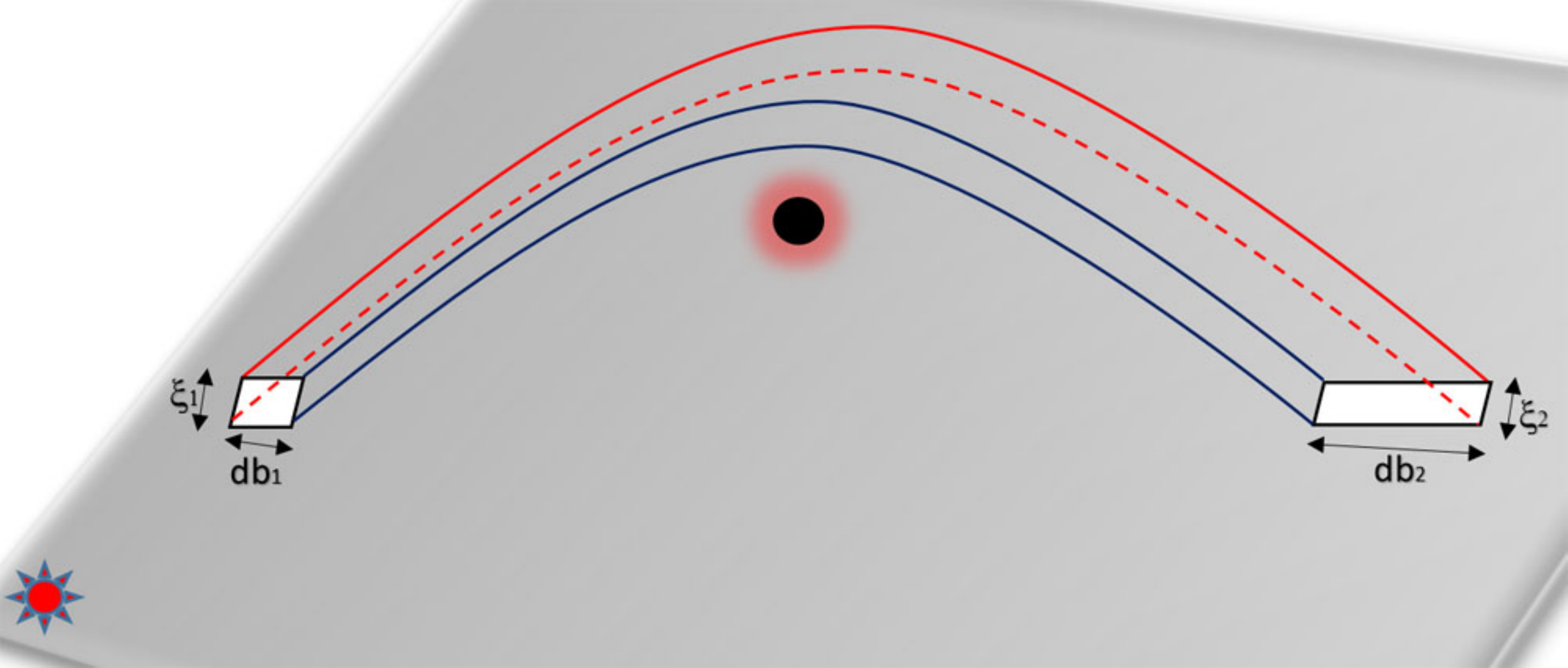}
\caption{Schematic illustration of the propagation of a thin
straight light beam in a spherically symmetric gravitational
field. The distance from the central object to the source is
assumed to be much smaller than the distance to the observer
(Earth); the height and width of the beam change as it propagates
in the gravitational field.} \label{R1}\end{figure}

 Consider a thin light beam from a distant star propagating in a gravitational field.
Suppose that the gravitational field is spherically symmetric and,
therefore, the light beam propagates near a single plane (in which
the source of the gravitational field lies). The beam parameters
before gravitational scattering will be denoted by index
"${{}_1}$"; after gravitational scattering, an observer on Earth
records the parameters of the starlight in the beam denoted by
index "${{}_2}$". Let the source be at a distance $l_1$ from the
gravitational center and the observer be on Earth (at a distance
$l_2$ from the gravitational center), with ${l_1 << l_2}$. For
convenience, we will confine the light beam to a rectangular
section, with the beam width in a direction parallel to the plane
under consideration being assumed to be equal to ${db}$ (see
Fig.~\ref{R1}). Let the extreme rays of light along the beam width
differ from one another by ${dh}$, where $h$ is the impact
parameter of the photons in the ray relative to the gravitational
center (this is an integral of motion for a photon
(see~\cite{Shatskiy2014}). Since the gravitational field changes
along the beam width, the width db changes due to the deviation of
null geodesic photons in a non-uniform gravitational field. The
beam width ${db_1}$ far from the gravitational center (before
scattering) must coincide with ${dh}$. We assume the beam
thickness in a direction orthogonal to the plane of the beam width
to be equal to $\xi$.

All null geodesics (rays of light) with identical impact
parameters converge at a single point after gravitational
deflection and thereafter again begin to diverge (see,
e.g.,~\cite{Zaharov1997}).

Let ${\varphi^{tot} (h)}$ be the total photon deflection
angle\footnote{The total photon deflection angle ${\varphi^{tot}
(h)>\pi}$ is measured from the field center relative to the light
source.}  (on the way from the source to the observer). In the
most general spherically symmetric metric dependent only on the
radial coordinate, this angle is defined by the expression
(see~\cite{Shatskiy2014}):
\begin{eqnarray}
\varphi^{tot} (h) = 2\int\limits_{r_{min}(h)}^\infty \frac{-h\,
dr}{R^2\sqrt{f_r(1/f_t - h^2/R^2)}} \, , \quad R^2(r_{min}) = h^2
f_t(r_{min}) \, . \label{varphi_h}\end{eqnarray} The metric
functions in this expression are defined by the metric
\begin{eqnarray}
ds^2=f_t(r)\, dt^2-\frac{dr^2}{f_r(r)} - R^2(r)\, d\Omega^2 \, ,
\quad d\Omega^2 := d\theta^2 + \sin^2\theta\, d\varphi^2 \, .
\label{metrkn}\end{eqnarray} For the Schwarzschild metric, for
example, we have: ${f_t = f_r = (1-r_g/r)}$, ${R^2=r^2}$.

Since we will be interested in strong gravitational scattering
(when the total change in light beam direction is greater than or
approximately equal to ${\sim \pi/2}$), the point at which the
rays again converge is at a distance ${l_{eins}<<l_2}$ from the
gravitating center. Consequently, the light beam parameters near
the Earth are defined by the relations\footnote{In Eqs.
(\ref{xi2}-\ref{db2}) we took into account the natural divergence
of the light beam under consideration, i.e., the increase in the
beam cross section ${dS}$ unrelated to the gravitational field;
this increase ${dS}$ must be inversely proportional to the square
of the distance from the star $l$, because ${\xi\propto 1/l}$ and
${db\propto 1/l}$ (these are the first terms on the right-hand
sides of Eqs. (\ref{xi2}-\ref{db2}). We also took into account the
proportion following from the relation for the light cones before
and after scattering: ${\xi_1/h = \Delta\xi_2/(l_2
\sin\,\varphi^{tot})}$.}:
\begin{eqnarray}
\xi_2 = | \xi_1\cdot(l_2/l_1) - l_2\, \xi_1\, (\sin\varphi^{tot})/h \label{xi2} | \\
db_2 = | db_1\cdot(l_2/l_1) + l_2\, \delta\varphi^{tot} | \approx
| dh\cdot(l_2/l_1) - l_2
\partial_h\varphi^{tot}\, dh | \label{db2}
\end{eqnarray}
Under weak gravitational scattering of a massive body, we have
${\partial_h\varphi^{tot}<0}$ and ${\sin\varphi^{tot}<0}$;
consequently, only the positive quantities are summed in Eqs.
(\ref{xi2}) and (\ref{db2}) for this case.

Since the total energy flux in the beam must be the same before
and after its scattering by the gravitating center, the apparent
brightness of light I in the beam must be inversely proportional
to the cross-sectional area of the beam ${dS:=\xi\, db}$, i.e.,
${I_2/I_1 = (\xi_1 db_1)/(\xi_2 db_2)}$.

Let ${I^{ordinary}_2}$ be the intensity of light in the beam near
the Earth that would be if there was no gravitational field:
\begin{eqnarray}
\frac{I^{ordinary}_2}{I_1} := \frac{l_1^2}{l_2^2}
\label{dI2}\end{eqnarray} This formula describes the natural
divergence of the beam and the natural brightness attenuation in
it.

Denote the coefficient of gravitational attenuation of the
intensity in the beam by ${\kappa := I_2 / I^{ordinary}_2}$. From
Eqs. (\ref{xi2}-\ref{dI2}) for $\kappa$ we then have
\begin{eqnarray}
\kappa = \frac{1}{|(1 - l_1\partial_h\varphi^{tot})(1 - l_1
\sin\varphi^{tot}/h)|} \label{kappa}\end{eqnarray} For the
Schwarzschild metric, at large $h$ (under weak gravitational
scattering according to \cite{LL2}, \S 101), we have
${\varphi^{tot}\approx \pi +2r_g/h}$, i.e., we obtain ${\kappa\to
1}$ at ${h>>r_g}$, as it must be.

Let us now estimate the strong gravitational scattering in the
Schwarzschild metric for the stars nearest to the central black
hole in our Galaxy. One of the stars nearest to the central black
hole, S2 (in Sagittarius; see~\cite{S2}), is at a distance of
${\approx 18\cdot 10^{9}km}$ from it ${(\approx 1500 r_g)}$ and
the attenuation coefficient for this star turns out to be
${\kappa_{S2}\approx 3\cdot 10^{-6}}$ (at
${\varphi^{tot}=1.5\pi}$).

Of course, an observer on Earth will see the source directly (with
a natural brightness attenuation approximately equal to ${\approx
l_1^2/l_2^2}$) and (hypothetically), after strong gravitational
scattering, from a ring with a radius of about ${3\sqrt{3}r_g/2}$
around the central black hole, where the source’s brightness near
the Earth is defined by Eqs. (\ref{dI2}) and (\ref{kappa}). If
there are quite a few such stars (at the very center of the Milky
Way), then the entire ring with a radius of about ${\sim
3\sqrt{3}r_g/2}$ around the black hole turns out to be shining.

A star like our Sun placed at the center of the Milky Way will be
seen as a star of approximately the 19th magnitude. Since a
brightness attenuation by a million times corresponds to ${\delta
m=15}$, the Sun placed at the center of the Milky Way must be seen
as a star of the 34th magnitude after the strong gravitational
scattering considered above. This is still an unattainable
sensitivity level for present-day telescopes. However, as is well
known, there exist stars that have an absolute luminosity higher
than the solar one by many orders of magnitude.

\section{GRAVITATIONAL SCATTERING BY A BLACK–WHITE HOLE}
\label{s3}

In contrast to ordinary (external) gravitational scattering by a
massive body, under internal gravitational scattering of light
from another universe, the greater the h, the greater the total
photon deflection angle\footnote{The passage of the ray with
${h=0}$ corresponds (by definition) to rectilinear propagation of
the ray of light in a wormhole.} ${\varphi^{tot}}$ on the way from
the source to the observer. Since ${\varphi^{tot}}$ can be zero
(at ${h=0}$) and the derivative ${\partial_h\varphi^{tot}}$ can be
positive, the light attenuation coefficient $\kappa_{wh}$ for a
wormhole or black–white hole can generally be arbitrary (large,
small, or of the order of unity)! This makes it possible to see
the light of stars from another universe and to distinguish it
from the light of stars in our Universe.

Consider the main properties of strong gravitational scattering by
a wormhole~\cite{Shatskiy2009} or black–white
hole~\cite{Shatskiy2014}. Each point source in the Universe from
the black hole will be seen as an infinite number of images in the
Universe from the white hole. In this case, each image corresponds
to its impact parameter $h$. The two main images of the source
correspond to the total photon deflection through angles
${\varphi^{tot}_1<2\pi}$ and ${\varphi^{tot}_{1'}=2\pi
-\varphi^{tot}_1}$, implying that the photons go around the center
on different sides (these two images will be seen on the same line
with the center and on different sides from it). The next two
images (with larger impact parameters $h$) correspond to the
photon deflection angles ${\varphi^{tot}_2=2\pi +
\varphi^{tot}_1}$ and ${\varphi^{tot}_{2'}=2\pi +
\varphi^{tot}_{1'}}$. The next pair is ${\varphi^{tot}_3=4\pi +
\varphi^{tot}_1}$ and ${\varphi^{tot}_{3'}=4\pi +
\varphi^{tot}_{1'}}$. And so on.

We considered two models: a Bronnikov–Ellis wormhole
(see~\cite{Ellis1973, Bronnikov1973}) and a Reissner–Nordstrom
black–white hole (see~\cite{Shatskiy2014}). For the model of a
Bronnikov–Ellis wormhole, the metric is written as
\begin{eqnarray}
ds^2 = (1-r_g/R)\, dt^2 - dr^2 - R^2\, d\Omega^2 ,\quad R^2:=
r^2+r_0^2 , \quad r_g < r_0 . \label{ds_wh1}\end{eqnarray} In this
model, for the maximum impact parameter of a photon that passes
from another universe into our Universe, we have ${h_{max} =
r_g\sqrt{27}/2}$ (at ${\frac{2}{3}r_0 < r_g < r_0}$) and ${h_{max}
= r_0^{3/2}/\sqrt{r_0-r_g}}$ (at ${r_g \le \frac{2}{3}r_0}$).

For ${r_g=0}$ and ${h\to 0}$, we have the asymptotics
${\varphi^{tot}(h)\to \pi h/r_0}$ and, therefore,
${\kappa_{wh}(h)\to r_0^2/(2\pi l_1^2)}$; in a more general case
${(r_g>0)}$, the asymptotics of ${\kappa_{wh}}$ does not depend on
$h$ either (see~\cite{Shatskiy2009}). For the model of a
Reissner–Nordstrom black–white hole,
\begin{eqnarray}
ds^2 = f(r)\, dt^2 - f^{-1}(r)\, dr^2 - r^2\, d\Omega^2 ,\quad
f(r):= \left(1-\frac{r_c}{r}\right)\left(1- \frac{r_h}{r}\right) .
\label{ds_wh2}\end{eqnarray} In this case, the charge of the black
hole is ${Q = \sqrt{r_c r_h}}$ and its mass is ${M = (r_c +
r_h)/2}$, where $r_c$ and $r_h$ are the inner and outer horizon
radii.

For the maximum impact parameter of a photon in this model that
passes from another universe into our Universe, we have ${h_{max}
= r_{m}^2/\sqrt{Q^2 - 2 M r_{m} + r_{m}^2}}$, where ${r_{m} :=
1.5M + 0.5\sqrt{9M^2-8Q^2}}$. Since we have the asymptotics:
${\varphi^{tot}(h)\to 2\sqrt{h/Q}}$, for ${\kappa_{wh}(h)\to \pi h
Q/(4 l_1^2)}$ -- see~\cite{Shatskiy2014}.

\begin{figure}
\includegraphics[width=0.95\textwidth]{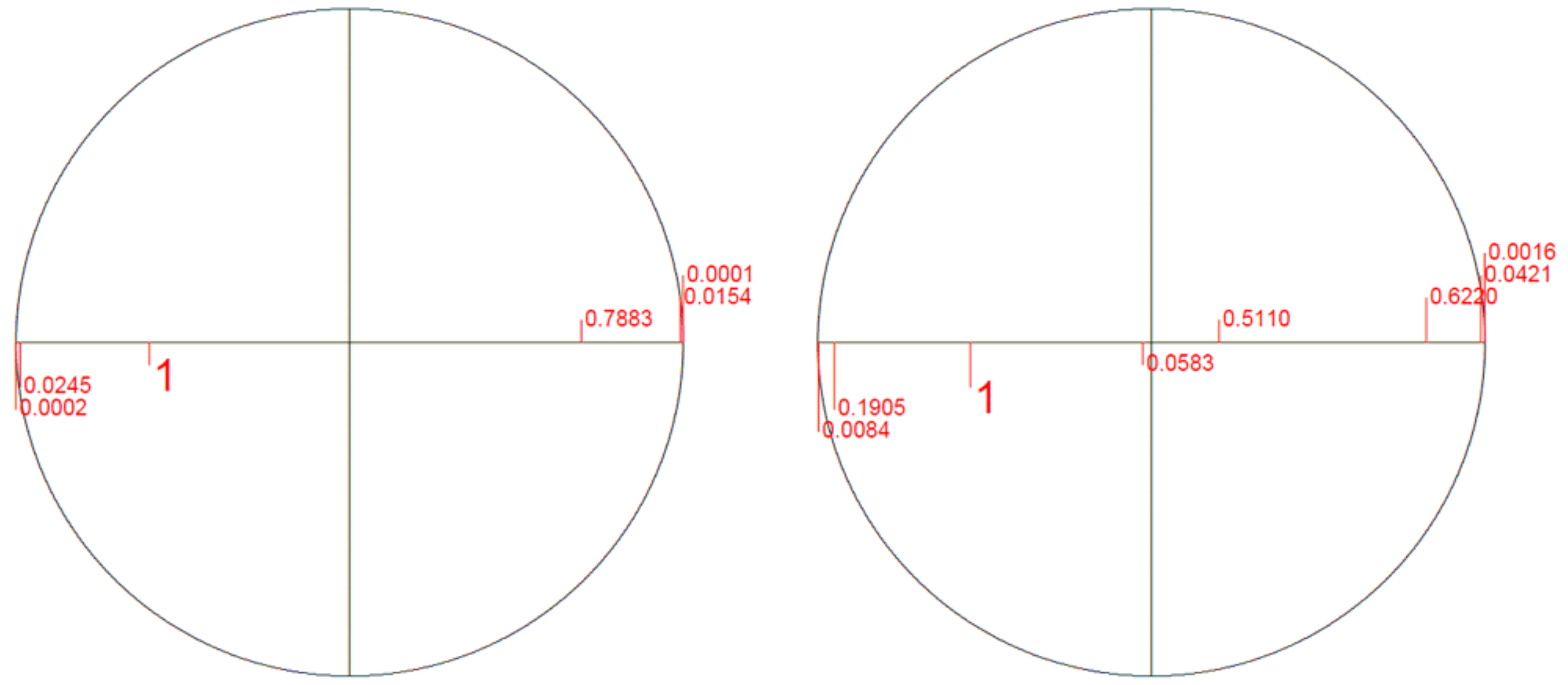}
\caption{System of images from a single point source in another
universe observed through a Bronnikov–Ellis worm hole (a) and a
Reissner–Nordstrom black–white hole (b). The circumference radii
are ${h_{max}}$. The wormhole parameters are ${r_g = 0.9 r_0}$ and
${\varphi^{tot}_1 = 0.9\pi}$. The black–white hole parameters are
${Q = 0.6 M}$ and ${\varphi^{tot}_1 = 0.5\pi}$.}
\label{R_bwh}\end{figure}

Figure~\ref{R_bwh} presents the dependences of the relative
brightness ${I_{rel}(h)}$ of the images\footnote{Relative to the
brightest image.} for a single star seen from another universe
through a wormhole and black–white hole. The vertical lines in
this figure mark the places of the first three (four) pairs of
visible virtual images, and the corresponding numbers mark the
relative brightnesses of these images. The brightnesses of all
these images generally turn out to be different; since the
brightness of the succeeding pairs of images decreases rapidly
starting from the third pair of images, it makes sense to consider
no more than the first two pairs of stellar images.

An important conclusion of this section is that all images of a
single star observed from another universe through a wormhole lie
on the same straight line, while their number (for the main images
comparable in brightness) is always greater than one. Therein also
lies the fundamental difference from strong gravitational
scattering by an ordinary black hole (when the ray of light always
remains in our Universe): there will be only one bright image,
while the brightness of the remaining images can already be
neglected.

\section{MODELING THE IMAGE VISIBILITY AMPLITUDE UNDER GRAVITATIONAL
SCATTERING IN A CENTRALLY SYMMETRIC GRAVITATIONAL FIELD}
\label{s4}

\begin{figure}
\includegraphics[width=0.95\textwidth]{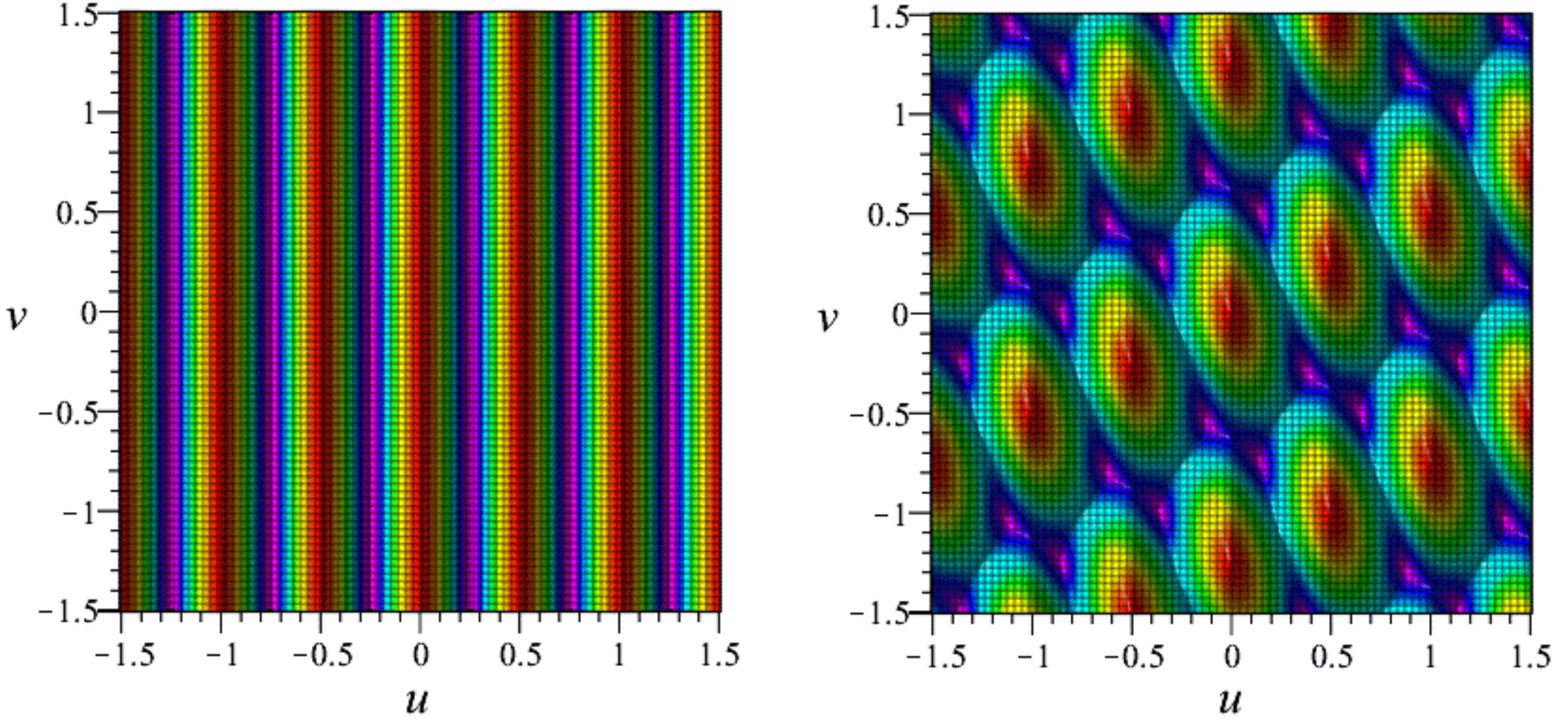}
\caption{Form of the CFD from several (different) point sources in
coordinates ${\rm (u, v)}$. (a) Two point sources with an angular
distance between the sources equal to ${2a}$ [rad]; the relative
brightness of the first and second sources are ${2}$ and ${3}$
arbitrary units, respectively (the latter is a factor of ${1.5}$
brighter). (b) Three point sources (not on a single straight
line); their angular coordinates ${(x_i, y_i)}$ and relative
brightnesses $I_i$ are: ${x_1 = a}$, ${y_1 = 0}$, ${I_1 = 3}$;
${x_2 = -a}$, ${y_2 = 0}$, ${I_2 = 3}$; ${x_3 = 0.6a}$, ${y_3 =
0.8a}$, ${I_3 = 4}$. The scale of the unit in coordinates ${\rm
(u,v)}$ corresponds to ${\lambda [sm]/(a[rad])}$.}
\label{R_M2_3}\end{figure}

Consider the various cases of modeling sources around a black
hole.

Individual point sources are modeled by delta functions:
${I_i(x,y)=A_i\, \delta(x-x_i)\, \delta(y-y_i)}$. For several
point sources, the CFD is then the absolute value of the sum of
their Fourier transforms:
\begin{eqnarray}
|V_i{\rm (u,v)}| = \left|\sum\limits_{i} A_i\exp\left[ -2\pi i{\rm
(x_i u+y_i v)}/\lambda \right] \right|
\label{3furye1}\end{eqnarray}

{\bf 1. The model of a single point source.} The CFD is a
constant: ${|V_1{\rm (u,v)}| = A_1 = const}$.

{\bf 2. The model of two point sources on a single straight line.}
The results for this model are displayed in Fig~\ref{R_M2_3}a. It
can be seen from these results that the point sources located on a
single straight line give a characteristic picture for the CFD:
the CFD in this case changes only along one direction while
remaining constant in an orthogonal direction. It can be shown
that this model does not change fundamentally if the number of
comparable (in brightness) point sources (on a single straight
line) will be greater than two\footnote{Such a situation is
typical for the passage of starlight through a wormhole or
black–white hole (several images with a comparable brightness from
a single point source lie on a single line;
see~\cite{Shatskiy2014}.}. In this case, the distance between the
sources (or the wormhole parameters) can be judged from the
periodicity of the CFD change.

\begin{figure}
\includegraphics[width=0.95\textwidth]{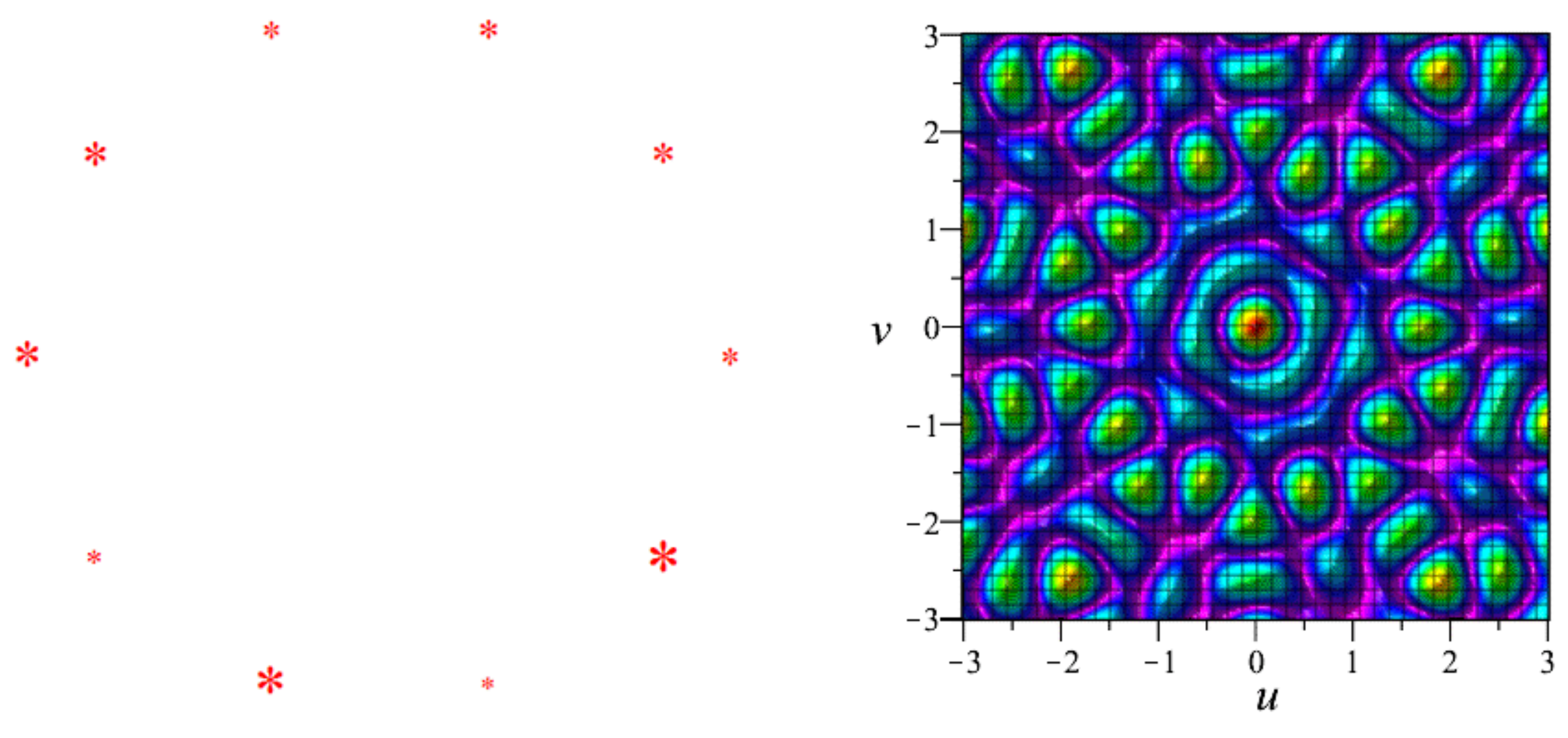}
\caption{(a) The model of ten comparable (in brightness) point
sources that lie on a single circumference with an angular radius
${R}$ [rad] at equal distances from one another. The relative
brightnesses of these sources are: ${I_1 = 1}$, ${I_2 = 0.9}$,
${I_3 = 0.8}$, ${I_4 = 1.1}$, ${I_5= 1.2}$, ${I_6 = 0.7}$, ${I_7 =
1.3}$, ${I_8 = 0.6}$, ${I_9 = 1.4}$ and ${I_{10} = 0.8}$. (b) The
CFD for this model. The scale of the unit in coordinates ${\rm
(u,v)}$ corresponds to ${\lambda [sm]/(R[rad])}$.}
\label{R_M4}\end{figure}

\begin{figure}
\includegraphics[width=0.95\textwidth]{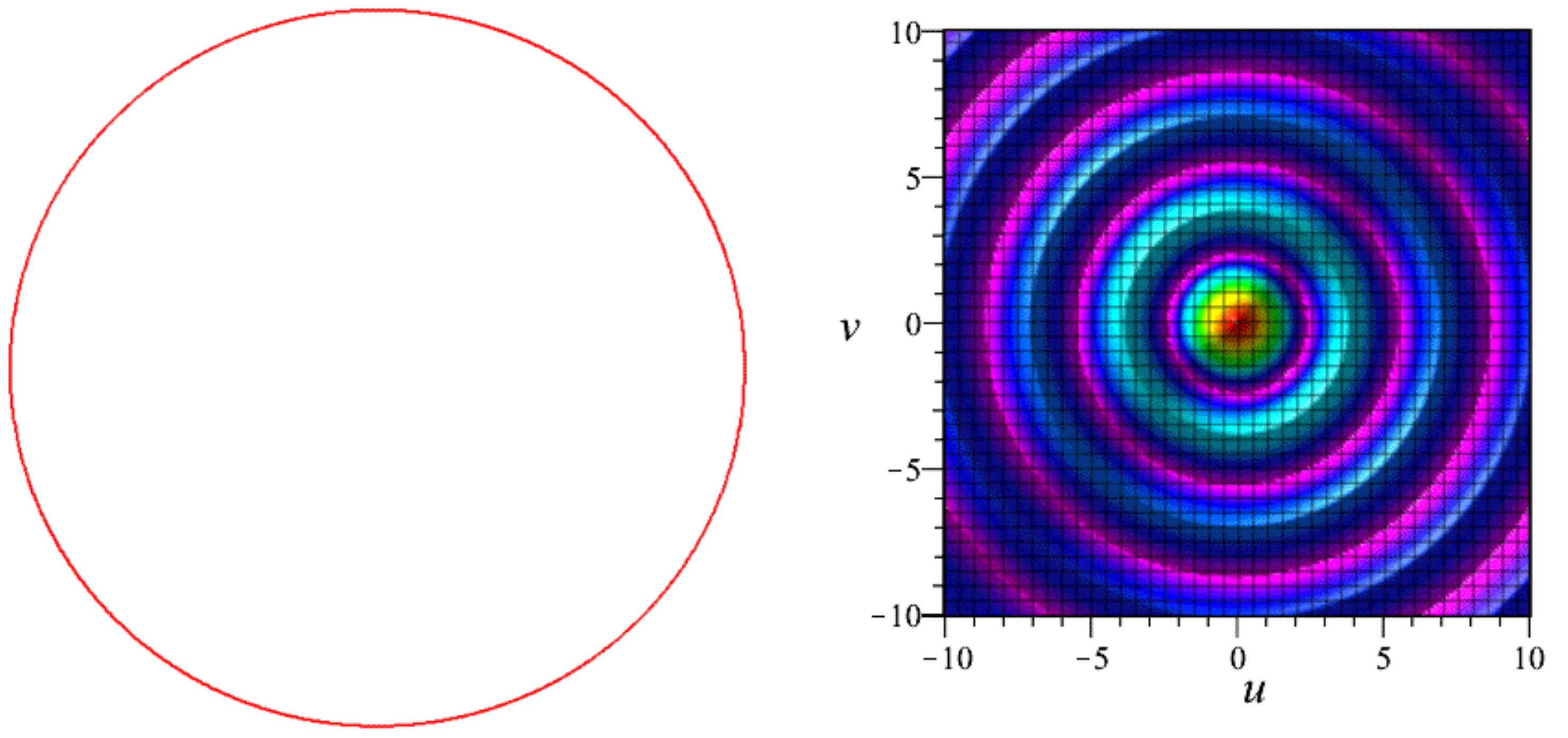}
\caption{(a) The model of a uniformly shining circumference with
an apparent radius ${r_{ring}}$ [rad]. (b) The CFD of the Fourier
transform for this model. The scale of the unit in coordinates
${\rm (u,v)}$ corresponds to ${lambda [sm]/(r_{ring}[rad])}$.}
\label{R_M5}\end{figure}

{\bf 3. The model of three point sources that do not lie on a
single straight line.} The CFD will change in both orthogonal
directions, and these changes will also be periodic
(see~\ref{R_M2_3}b). Again, the mutual distance between the
sources can be judged from these periods.

{\bf 4. The model of ten comparable (in brightness) point sources
that lie on a single circumference at equal distances from one
another.} The CFD for this model is shown in Fig.~\ref{R_M4}.
According to the form of this CFD, it does not matter any longer
whether these sources lie on a single circumference or not (at
such a number of comparable (in brightness) and equidistant point
sources).

{\bf 5. The model of a uniformly shining circumference.} The CFD
of the Fourier transform for this model is
\begin{eqnarray}
|V_{ring}{\rm (u,v)}| = const\cdot |J_0(\eta r_{ring})|  \, ,
\quad \eta := \sqrt{{\rm u^2+v^2}}/\lambda \, .
\label{3furye2}\end{eqnarray} Here ${J_p(x)}$ -- is a Bessel
function of order "$p$"\, (see.~Fig.~\ref{R_M5}).

\begin{figure}
\includegraphics[width=0.95\textwidth]{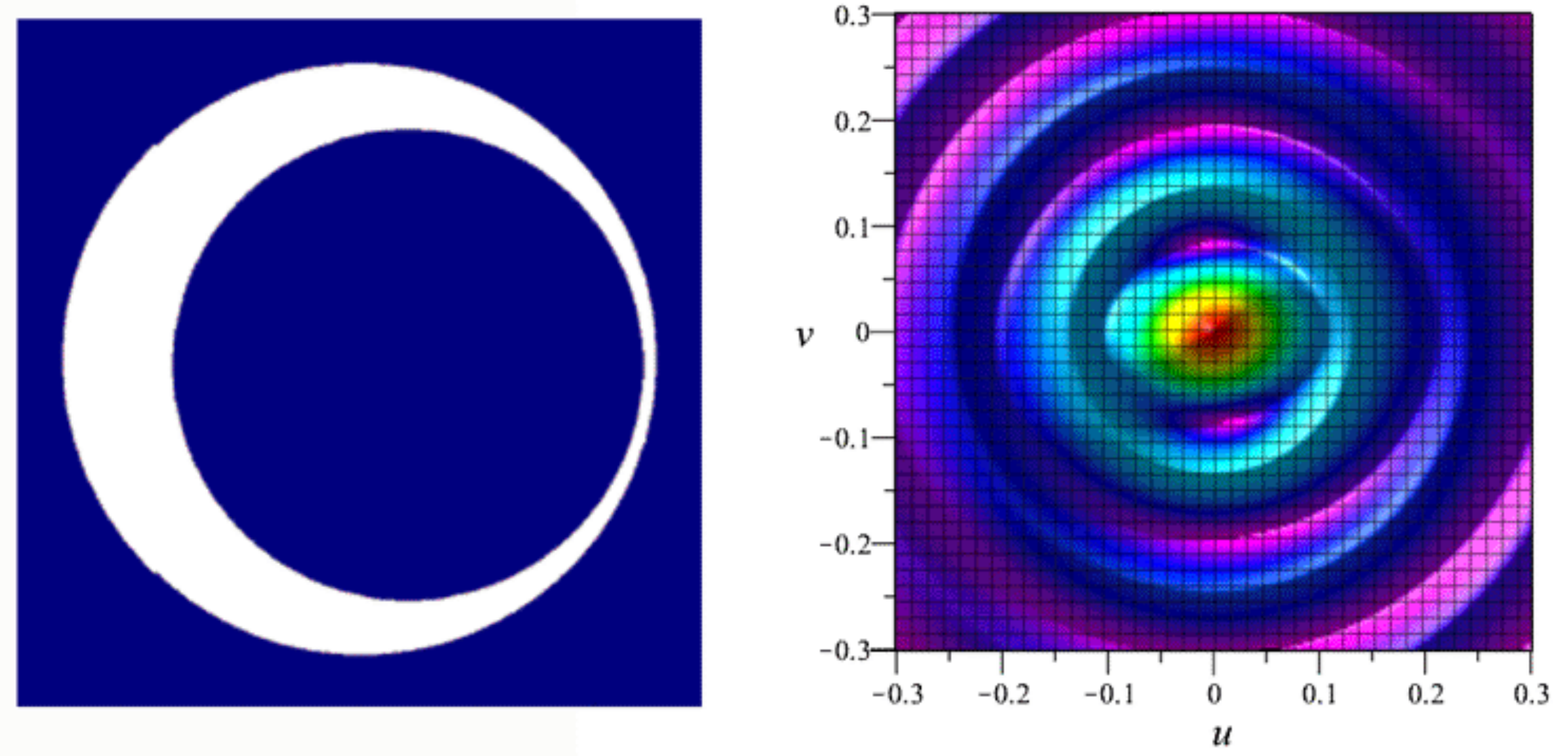}
\caption{(a) The crescent model (simulates the model of a
nonuniformly shining circumference): inside a uniformly shining
disk with an apparent diameter ${r_{out}}$ [rad] there is a dark
disk with a smaller diameter displaced from its center;
${r_{in}=0.8 r_{out}}$, ${x_{c}= 0.16 r_{out}}$ and ${y_{c}= 0.02
r_{out}}$. (b) The CFD of the Fourier transform for this model.
The scale of the unit in coordinates ${\rm (u,v)}$ corresponds to
${\lambda [sm]/(r_{out}[rad])}$.} \label{R_M6}\end{figure}

\begin{figure}
\includegraphics[width=0.95\textwidth]{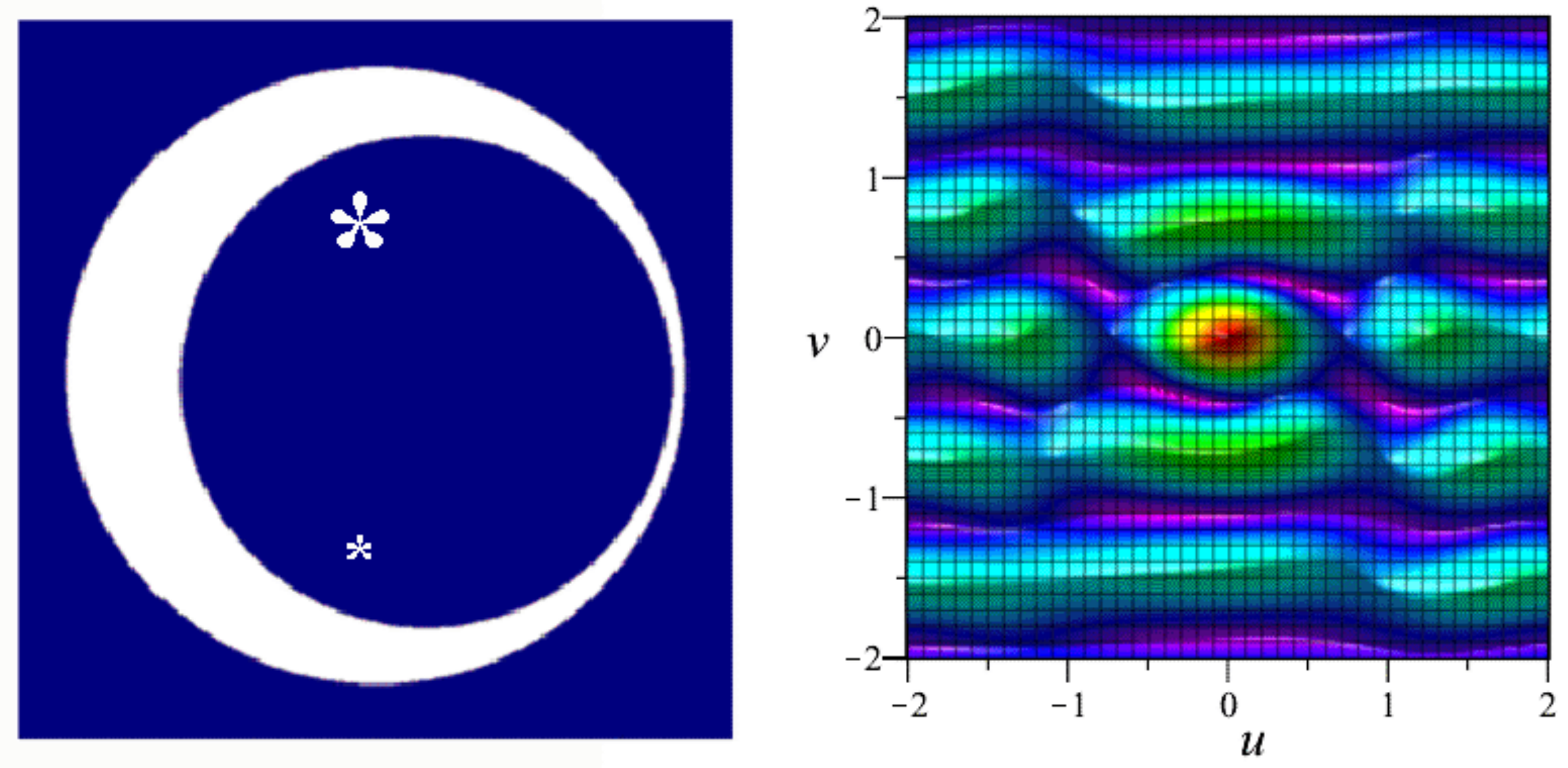}
\caption{(a) The crescent model, ${r_{in}=0.8 r_{out}}$ (see
Fig.~\ref{R_M6}), with weight ${A_0=1}$ (and a corresponding
brightness ${\approx 1.13 r_{out}^2}$) plus two comparable (in
brightness) point sources with weights ${A_1=0.1}$, ${A_2=0.05}$
and coordinates ${y_1 = 0.6 r_{out}}$, ${y_2 = -0.7 r_{out}}$
(see~(\ref{3furye4})). (b) The CFD of the Fourier transform for
this model. The scale of the unit in coordinates ${\rm (u,v)}$
corresponds to ${\lambda [sm]/(r_{out}[rad])}$.}
\label{R_M7}\end{figure}

{\bf 6. The crescent model (simulates the model of a non-uniformly
shining circumference).} Inside a uniformly shining disk with a
diameter ${r_{out}}$ there is a dark disk with a smaller diameter
${r_{in}}$ displaced from its center (see~\cite{crescent}):
\begin{eqnarray}
|V_{crescent}{\rm (u,v)}| = \frac{const}{\eta}\cdot \left|
r_{out}\, J_1(\eta r_{out}) - e^{-2\pi i {\rm (x_c u + y_c
v)}/\lambda}\, r_{in}\, J_1(\eta r_{in}) \right|
\label{3furye3}\end{eqnarray} Here, $x_c$ and $y_c$ are the
coordinates of the displacement of the inner disk center relative
to the outer disk center (see Fig.~\ref{R_M6}).

{\bf 7. The crescent model plus two comparable (in brightness)
point sources (see Fig.~\ref{R_M7}).} We have
\begin{eqnarray}
|V_{crescent+points}{\rm (u,v)}| = \left|\frac{A_0}{\eta}\left[
r_{out}\, J_1(\eta r_{out}) - e^{-2\pi i {\rm (x_c u+y_c
v)}/\lambda}\, r_{in}\, J_1(\eta r_{in}) \right] + \sum\limits_{j}
A_j e^{-2\pi i (y_j {\rm v})/\lambda} \right| \quad
\label{3furye4}\end{eqnarray} It is important to compare the
relative brightnesses of the crescent and the point source in this
model. Therefore, we will assume that a unit brightness, the
brightness of a ${\rm 1x1}$ area with a unit weight, corresponds
to the delta function (i.e., a point source with a unit weight).
The brightness of the crescent with a unit weight will then be
${\pi (r_{out}^2-r_{in}^2)}$.

\begin{figure}
\includegraphics[width=0.95\textwidth]{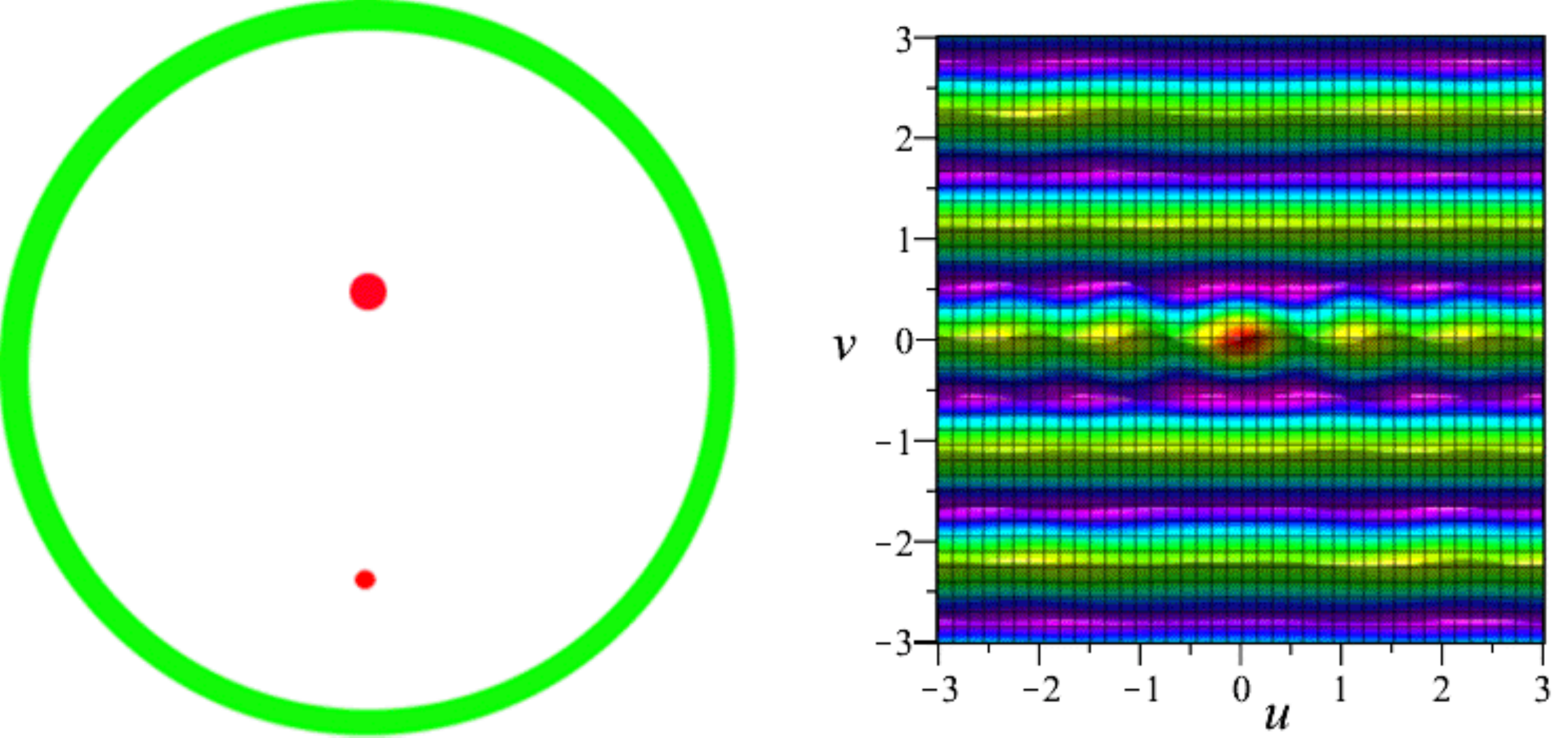}
\caption{(a) The model of a thick circumference plus two
oppositely displaced point sources with the same total brightness
${I_{ring} = 1}$. The coordinates of the sources are ${y_1=+0.2
r_{out}}$ and ${y_2=-0.7 r_{out}}$; the relative brightnesses of
the sources are ${I_1 = 0.8}$ and ${I_2 = 0.2}$. (b) The CFD for
this model. The scale of the unit in coordinates ${\rm (u,v)}$
corresponds to ${\lambda [sm]/(r_{out}[rad])}$.}
\label{R_M8}\end{figure}

{\bf 8. The model of a thick circumference plus two oppositely
displaced point sources with the same total brightness} (see
Fig.~\ref{R_M8}):
\begin{eqnarray}
|V_{ring+points}{\rm (u,v)}| = A_1 \left|\frac{r_{out}\, J_1(\eta
r_{out}) - r_{in}\, J_1(\eta r_{in})}{\eta\, r_{out}^2} + I_1
e^{-2\pi i y_1 {\rm v}/\lambda} + (1-I_1) e^{-2\pi i y_2 {\rm
v}/\lambda} \right| \quad \label{3furye5}\end{eqnarray} The
brightness of the thick circumference and the total brightness of
both point sources in the model are chosen to be the same. For
this purpose, we assume that ${r_{in} := r_{out}\sqrt{1-1/\pi}}$
and ${I_1\in [0,\, 1]}$.

In this model, it is also important to determine the minimum
ratios ${y_1/r_{out}}$ and ${y_2/r_{out}}$ at which the visual
asymmetry of the picture is still seen.

\section{Asymmetry of the Correlated Flux Density}
\label{s_ass}

As has become clear from the previous section, one of the
characteristic features in the CFD when wormholes or black–white
holes are observed can be its asymmetry in coordinates ${\rm
(u,v)}$. Therefore, it makes sense to introduce a measure of this
asymmetry. Suppose that the model under consideration is
completely symmetric with the center of symmetry at the coordinate
origin on the ${\rm (u,v)}$-plane. For any two rays on this plane
originating from the coordinate origin, the CFD will then be
identical at equal distances from the center. The asymmetry in
this case is zero, as, for example, in models 1 and 5 considered
in the previous section.

Let us define the degree of asymmetry As as
\begin{eqnarray}
As :=
\frac{max\{Int(\tau)\}-min\{Int(\tau)\}}{max\{Int(\tau)\}+min\{Int(\tau)\}}
,\quad Int(\tau) := \int\limits_{0}^{+\infty} \left|
\frac{d|V(\eta , \tau)|}{d\eta} \right|\, d\eta .
\label{As1}\end{eqnarray} Here, we have introduced the polar
coordinate ${(\eta , \tau)}$ instead of the Cartesian coordinates
${\rm (u,v)}$:
\begin{eqnarray}
{\rm u}(\eta , \tau) := \lambda\eta\cdot\cos(\tau) ,\quad {\rm
v}(\eta , \tau) := \lambda\eta\cdot\sin(\tau) . \label{As2}
\end{eqnarray}
For this definition, the maximum possible degree of asymmetry is
equal to one: ${max\{As\} = 1}$. The degree of asymmetry for
models 2, 3, 4, 6, 7, and 8 lies within the range ${0<As<1}$. We
numerically calculated the asymmetry for two of these models:

Model 6: ${As^6\approx 0.47}$ at ${x_c=0.2r_{out}}$, ${y_c=0}$ and
${As^6\approx 0.36}$ at ${x_c=0.1r_{out}}$;

Model 8: ${As^8\approx 0.13}$ at ${y_1 = +0.2r_{out}}$, ${y_2 =
-0.7r_{out}}$ and ${a_1 = 0.8}$.

For models 6 and 8, we took a finite range of
integration\footnote{Note that ${Int(\tau)}$ does not depend on
the units of measurement of $\lambda$, because the scale factor
cancels out after integration.}, ${\eta\in [0,\, 10]}$, when
calculating the asymmetry in integral (\ref{As1}).

As can be seen from these data, model 6 possesses an even greater
asymmetry than model 8 for such a definition. However, model 6 is
also used to describe the shadow from the black hole
(see~\cite{crescent}). Therefore, it is impossible to
unambiguously distinguish the effects related to non-central point
sources from other effects (related, for example, to ordinary
black holes) based only on the asymmetry (\ref{As1}) in the range
of integration ${\eta\in [0,\, 10]}$.

To improve our method, let us modify Eq. (\ref{As1}) in such a way
that the analogous effects from other objects (for example, from
ordinary black holes) are excluded in the asymmetry determination.
For this purpose, notice that only the CFD oscillations associated
with the pair point sources that lie on a single straight line
with the coordinate origin make a major contribution to the
asymmetry at great distances from the coordinate origin. This is
just what we need to describe the pair of images from a single
point source when observed through a wormhole or black–white hole.
Therefore, to reveal the sought-for effects (precisely from such
point sources) in the asymmetry, let us modify one of Eqs.
(\ref{As1}):
\begin{eqnarray}
Int_{mod}(\tau) := \int\limits_{\eta_1}^{\eta_2} \left|
\frac{d|V(\eta , \tau)|}{d\eta} \right|\, d\eta ,
\label{As2}\end{eqnarray} where ${1 << (\eta_2 - \eta_1) << \eta_1
< \eta_2}$. In such a modified case, we obtain the following
values for models 6 and 8 for the parameters ${\eta_1 = 100}$ and
${\eta_2 = 110}$:

model 6: ${As_{mod}^6\approx 0.30}$ at ${x_c=0.2r_{out}}$,
${y_c=0}$ and also at ${x_c=0.1r_{out}}$;

model 8: ${As_{mod}^8\approx 0.93}$ at ${y_1 = +0.2r_{out}}$,
${y_2 = -0.7r_{out}}$ and ${a_1 = 0.8}$.

Thus, the asymmetry in the modified definition (\ref{As2})
dominates and approaches unity in the cases of interest to us:
when pair point sources that lie on a single straight line with
the coordinate origin are present in the image. Of course, the
specific value of ${As_{mod}}$ depends on the choice of ${\eta_1 <
\eta_2}$, but we are interested only in the closeness of
${As_{mod}}$ to unity to distinguish the different possible
topologies of the observed object.

\section{DISCUSSION}
\label{s5}

It is clear from what has been said above that unambiguous
conclusions cannot always be drawn from the form of the CFD
distribution function alone.

For example, it will be difficult to determine in practice whether
the numerous point sources are located on a circumference or they
are distributed inside it from the form of the CFD distribution
(in Fig.~\ref{R_M4}).

The crescent model considered above can also correspond to
different physical situations: for example, the model of a
non-uniformly shining circumference (see Fig.~\ref{R_M6}) or the
model of an accretion disk that is shadowed by a black hole on one
side. In this case, the “blurred” crescent model (to allow for the
interstellar scattering effects; see, e.g.,~\cite{crescent}) is
used for the best agreement with the observations. Various kinds
of blurring are also often applied in other models. All of this
makes numerous simulations of such a kind difficult to distinguish
from one another.

However, the models where there are distributions of sources along
a single straight line (see Figs.~\ref{R_M2_3}, \ref{R_M7} and
\ref{R_M8})) constitute an important exception from the aforesaid.
In these cases, a characteristic feature of such models is the
existence of a direction on the ${\rm (u, v)}$ plane in which the
CFD remains constant (at great distances from the center). It is
these cases that are particularly interesting for the detection of
wormholes and black–white holes (see,~\cite{Shatskiy2014}).

The combination (superposition) of models 2 and 6, whose results
are presented in Fig.~\ref{R_M7}, is some intermediate case (model
7). Signatures that will more likely be typical of a wormhole or
black–white hole can also be revealed here. It is important that a
periodicity in coordinate ${\rm v}$ (in the absence of a distinct
periodicity in coordinate ${\rm u}$) is still clearly visible even
at a crescent brightness that is greater than the total brightness
of the point sources by approximately seven and a half times.
Thus, this model has a stable signature of a black–white hole or
wormhole. This signature manifests itself more and more as one
recedes from the center, because the CFD of the crescent (or a
different brightness distribution symmetric in angle $\tau$)
decreases as ${\propto \eta^{-3/2}}$, while the CFD of pair point
sources with a comparable brightness oscillates with a constant
amplitude and, as a result, its contribution becomes dominant.

Model 8 shows that the presence of pair sources that lie on a
single straight line with the coordinate origin in the image
changes radically the situation. Such a model (just as model 7)
will possess an asymmetry close to unity! Detecting this signature
will allow one to talk about the discovery of an object like a
wormhole or black–white hole. This will be possible if the angular
distance between the pair sources in the model will be less than
or of the order of the angular size of the horizon diameter for
the corresponding black hole. However, the presence of a
sufficiently bright source in another universe whose photons
(coming to us) would produce a system of pair images (see
Fig.~\ref{R_bwh}) is also needed for this condition to be
realized. In this case, the pairs of such images will satisfy the
condition corresponding to Fig.~\ref{R_M8} and ${As_{mod}\approx
1}$.

Note that in this paper we have considered the case of a
monochromatic radiation spectrum and a certain relation between
the flux densities from different image details (for more details,
see Section~\ref{s4}). The CFD form that we presented for various
characteristic cases can vary as these assumptions change. The
models considered here are directly applicable only in the case of
a sufficiently high telescope resolution in recorded spectral band
width ${\Delta\lambda/\lambda}$. This condition,
${\Delta\lambda/\lambda<<1}$, is met for VLBI systems. However,
${\Delta\lambda/\lambda \sim 0.2}$ for optical and infrared
interferometers. Therefore, an assumption about the radiation
spectra of image details in the detected band ${\Delta\lambda}$
should be made to apply our models in the optical and infrared
ranges.

\end{document}